\documentstyle[11pt]{article}
\newcommand{\CS}[1]{&{#1}&}
\newcommand{\NN}{\nonumber}
\newcommand{\SIN}[1]{\sin\left({#1}\right)}

\newcommand{\ket}[1]{|{#1}\rangle}
\newcommand{\half}{{1\over 2}}

\newcommand{\BE}{\begin{equation}}
\newcommand{\EE}{\end{equation}}
\newcommand{\BEA}{\begin{eqnarray}}
\newcommand{\EEA}{\end{eqnarray}}
\newcommand{\BEAN}{\begin{eqnarray*}}
\newcommand{\EEAN}{\end{eqnarray*}}
\newcommand{\TD}[1]{{\partial #1 \over \partial t}}
\newcommand{\LR}{\left(} \newcommand{\RR}{\right)}
\newcommand{\LS}{\left[} \newcommand{\RS}{\right]}
\newcommand{\LC}{\left\{ } 
\newcommand{\OP}{\Omega^{(+)}} \newcommand{\OM}{\Omega^{(-)}}
\newcommand{\OPM}{\Omega^{(\pm)}}

\newcommand{\HALF}[1]{{{#1}\over 2}}
\newcommand{\JETP}[1]{{\it Sov.~Phys.}~JETP~{\bf #1}}
\newcommand{\JSP}[1]{{\it J.~Stat.~Phys.}~{\bf #1}}
\newcommand{\PHYSICA}[1]{{\it Physica}~{\bf #1}}
\newcommand{\APNY}[2]{{\it Ann.~Phys.}~(NY)~{#1}~{#2}}
\newcommand{\EPL}[2]{{\it Europhys.~Lett.}~{#1}~{#2}}
\newcommand{\JP}[2]{{\it J.~Phys.}~{#1}~{#2}}
\newcommand{\JPSJ}[2]{{\it J.~Phys.~Soc.~Jpn.}~{#1}~{#2}}
\newcommand{\NP}[2]{{\it Nucl.~Phys.}~{#1}~{#2}}
\newcommand{\PL}[2]{{\it Phys.~Lett.}~{#1}~{#2}}
\newcommand{\PR}[2]{{\it Phys.~Rev.}~{#1}~{#2}}
\newcommand{\SSC}[2]{{\it Solid~State~Comm.}~{#1}~{#2}}

\begin{document}
\begin{center}
\vspace*{1.0cm}
\large{\bf Stochastic model and equivalent ferromagnetic spin
  chain with alternation} \\
\vspace*{1cm}
{\large {Akira Fujii
  \footnote{Present address: Universit{\"a}t zu K{\"o}ln, Institute
    f{\"u}r Theoretische Physik, Z{\"u}lpicher Str. 77, 50937 Germany}
  }} 
\vskip 0.5cm
{\normalsize 
{\sl Physikalisches Institut} \\
{\sl Universit{\" a}t Bonn} \\
{\sl Nussallee 12, 53115 Bonn} \\
{\sl GERMANY.  }
}
\vspace{1 cm}
\begin{abstract}
  We investigate a non-equilibrium reaction-diffusion model and
  equivalent ferromagnetic spin 1/2 $XY$ spin chain with alternating
  coupling constant.  The exact energy spectrum and the $n$-point hole
  correlations are considered with the help of the Jordan-Wigner
  fermionization and the inter-particle distribution function method.
  Although the Hamiltonian has no explicit translational symmetry, the
  translational invariance is recovered after long time due to the
  diffusion.  We see the scaling relations for the concentration and
  the two-point function in finite size analysis. 
\end{abstract}
\vspace{1cm}
\end{center}
\newpage 
\baselineskip=12pt
The study of the systems with reaction and diffusion has been an
attracting problem.  Especially, methods to determine the correlation
functions and to observe the approach to the thermal equilibrium in
one-dimensional models have been much developed recently. 
One of the recent success in this field is to obtain the exact
correlation functions including the coagulation term by the so-called
inter-particle distribution function (IPDF) method \cite{kadanoff,ann}.
If we appropriately tune the coupling constants, ({\it i.e.} the
reaction-diffusion {\it rates}) in this approach, the master equation
can be solved exactly.

On the other hand, the (partially) antiferromagnetic $XY$ spin chain
with alternating coupling constant has been investigated as a toy
model to consider the spin-Peierls phase transition or the Haldane
gap. The thermal equilibrium model with the alternation has been
studied by some authors \cite{pincus,okamoto}.  They have obtained the
energy spectrum with the mass gap proportional to the strength of the
alternation, the dimer correlation functions and so on. It has been
also shown that the dimerizing process lowers the ground state energy.
From those results, they have concluded that the uniform
antiferromagnetic $XY$-chain is unstable with respect to the
distortion.

In this paper, we investigate a thermal non-equilibrium model with
alternating coupling and diffusion.  It can be shown that the
stochastic model and the ferromagnetic $XY$ chain with the alternation
are mapped on each other by linear transformations\cite{ann,bonn}.
The basic tools to study these systems in this paper are the
Jordan-Wigner method and the IPDF method.  By the former method, which
can be used in both the ferromagnetic and (partially)
antiferromagnetic regimes, we can diagonalize the Hamiltonian exactly
and calculate the energy spectrum. By means of the latter, which is
available only in the ferromagnetic regime of the model, we obtain the
exact forms of the energy spectrum and the correlation functions.  Of
course, the overlapped results obtained in the ferromagnetic region by
both methods are the same.  Since in the ferromagnetic case, the
ground state occurs when all the spins are aligned, the time evolution
of the correlation functions is our main interest in this paper.

Hereafter, we investigate a spin-$1/2$ $XY$-model in a magnetic field
along the $z$-axis with alternation whose Hamiltonian is defined with
the alternation parameter $\Delta$ and the diffusion parameter
$\eta~(\geq 1)$ by \BEA H\CS{=}\sum_{i=1}^{L}\half\LS
1+(-1)^{i}\Delta\RS H_{i,i+1}, \NN \\ H_{i,i+1}\CS{=}-\eta\LR\eta
s^{x}_{i}s^{x}_{i+1}+{1\over\eta}s^{y}_{i}
s^{y}_{i+1}+s^{z}_{i}+s^{z}_{i+1}-\eta-{1\over\eta}\RR.
\label{eq:spinhmt} \EEA We assume that the length $L$ of the
spin-chain is an even integer and that periodic boundary conditions
($\vec{s}_{L+1}=\vec{s}_{1}$) are imposed.

Let us consider the exact diagonalization of the Hamiltonian
(\ref{eq:spinhmt}) by the Jordan-Wigner transformation
\cite{ann,bonn,barouch,hinrichsen}.  It should be mentioned that this
method is also available in both the antiferromagnetic ($\Delta<-1$)
and ferro-antiferromagnetic coexisting ($\Delta>1$) cases as well as
in the ferromagnetic ($|\Delta|<1$) one.  Due to alternation, the
unique band in the uniform model is split into two separated bands
with an energy gap.
Therefore, we introduce two kinds of fermions $b_{2n-1}$ and
$c_{n}$ corresponding to spins on the odd and even sites respectively.
The spin operators can be represented by these fermions as
\BEA
&&s^{+}_{2n-1}=b_{2n-1}^{\dagger}\exp\LR
  i\pi\sum_{j=1}^{n-1} b_{2j-1}^{\dagger}b_{2j-1}
  +i\pi\sum_{j=1}^{n-1}c_{2j}^{\dagger}c_{2j}\RR,\;
  s^{-}_{2n-1}=(s^{+}_{2n-1})^{\dagger}, \NN\\ 
 &&s^{+}_{2n}=c_{2n}^{\dagger}\exp\LR i\pi\sum_{j=1}^{n}
  b_{2j-1}^{\dagger}b_{2j-1}
  +i\pi\sum_{j=1}^{n-1}c_{2j}^{\dagger}c_{2j}\RR,\;
  s^{-}_{2n}=(s^{+}_{2n})^{\dagger}, 
\EEA 
where the integer $n$ runs from
$1$ to $L/2$.  In terms of the fermions $b$ and $c$, the boundary
conditions are given by $b_{L+i}=\pm b_{i}$ and $c_{L+i}=\pm c_{i}$
depending on whether ${\cal
  N}=\sum_{j=1}^{L/2}(b_{2j-1}^{\dagger}b_{2j-1}+
c_{2j}^{\dagger}c_{2j})$ is odd or even. Performing the Fourier
transformation \BE b_{2n-1}=\sqrt{2\over L}\sum_{k}e^{-{2\pi ik\over
    L}(2n-1)}{\bar b}_{k}, \quad c_{2n}=\sqrt{2\over
  L}\sum_{k}e^{-{2\pi ik\over L}\cdot 2n}{\bar c}_{k}, \EE and the
Bogoliubov transformation \BE ({\tilde b}_{k},{\tilde b}^{k},{\tilde
  c}_{k},{\tilde c}^{k})= ({\bar b}_{k},{\bar b}_{-k}^{\dagger},{\bar
  c}_{k},{\bar c}_{-k}^{\dagger}) S_{k}, \EE where $S_{k}$ is a
certain $(4\times 4)$-matrix, we can rewrite the Hamiltonian
(\ref{eq:spinhmt}) as 
\BEA 
&&H=\sum_{k}{\tilde H}_{k}, \NN \\ 
&&{\tilde
  H}_{k}+{\tilde H}_{\HALF{L}-k} = \lambda_{k}^{(+)} \LR{\tilde
  b}_{k}^{\dagger}{\tilde b}_{k} -{\tilde b}^{k\dagger}{\tilde
  b}^{k}\RR+ \lambda_{k}^{(-)} \LR{\tilde c}_{k}^{\dagger}{\tilde
  c}_{k} -{\tilde c}^{k \dagger}{\tilde c}^{k} \RR.  \EEA 
The energy
eigenvalues $\lambda_{k}^{(\pm)}$ are given by \BE\hspace{-23mm}
\lambda_{k}^{(\pm)}=-(\eta^{2}+1)\pm 2\eta\sqrt{\cos^{2}\LR{2\pi
    k\over L}\RR +\Delta^{2}\LS\sin^{2}\LR{2\pi k\over L}\RR
  +\LR{\eta-\eta^{-1}\over 2}\RR^{2}\RS}, \label{eq:jwspectrum} \EE
where the value of momentum $k$ depends on ${\cal N}$ as 
\BEA
  k\CS{=}0,1,2,\cdots,\HALF{L}-1,\quad{\rm for}~{\cal N}={\rm odd}, \NN\\ 
  k\CS{=}\HALF{1},\HALF{3},\cdots,\HALF{L}-\HALF{1},
\quad{\rm for}~{\cal N}={\rm
    even}.  
\EEA
Therefore, the ground state is given by the
half-filled state in this basis like \BE \hspace{-10mm}
\LR\prod_{k}{\tilde b}_{k}^{\dagger}\RR\LR\prod_{k}{\tilde
  c}_{k}^{\dagger}\RR \ket{\widetilde{vac}} \quad {\rm with}\quad
{\tilde b}_{k}\ket{\widetilde{vac}}= {\tilde
  b}^{k}\ket{\widetilde{vac}}= {\tilde c}_{k}\ket{\widetilde{vac}}=
{\tilde c}^{k}\ket{\widetilde{vac}}=0 \EE and the energy spectrum has
a gap 
$2\eta\Delta(\eta+\eta^{-1})$ at $2\pi k/L\sim \pi/2$ provided
$\Delta\neq 0$ as is shown for that without the diffusion ({\it i.e.}
the magnetic field) \cite{pincus,okamoto}. Contrary to the
antiferromagnetic case, the energy gap does not appear near the Fermi
surface in the ferromagnetic case.  It follows that the ground state
energy remains a monotonically increasing function of the alternation 
parameter 
$\Delta$ even if we add the
elastic energy of distortion to the Hamiltonian. 
It means that the instability due to the dimerization
cannot be observed in the ferromagnetic case.

On the other hand, as is shown in \cite{ann,bonn}, the spin
Hamiltonian (\ref{eq:spinhmt}) can be mapped onto a stochastic model
including coagulation and decoagulation processes by regarding the
spin-up and spin-down state as sites occupied by a particle (denoted
$A$) and empty (denoted $\phi$), respectively.  If the $i$-th site is
occupied (empty), we label $\sigma_{i}=1$ (0).  The stochastic model
to be considered hereafter is defined by the transition rates
$w_{\alpha\beta}(\mu,\nu)$ ($\alpha$, $\beta$, $\mu$, $\nu$ $\in$
${\bf Z}_{2}$) and arbitrary positive constants $a_{\pm}$.  Only the
following processes are allowed.
\begin{enumerate}
\item Diffusion \\ $A+\phi\leftrightarrow\phi+A$ at the rate
  $a_{\pm}w_{11}(01)=a_{\pm}w_{11}(10)$.
\item Coagulation \\ $A+A\rightarrow A+\phi$ and
  $A+A\rightarrow\phi+A$ at the same rate
  $a_{\pm}w_{01}(10)=a_{\pm}w_{10}(01)$.
\item Decoagulation \\ $A+\phi\rightarrow A+A$ and $\phi+A\rightarrow
  A+A$ at the same rate $a_{\pm}w_{01}(11)=a_{\pm}w_{10}(11)$.
\end{enumerate}
The above rates should be interpreted as below.  For example, let us
assume that the $i$-th and $(i+1)$-th sites are occupied and empty,
respectively, and that $i$ is an even (odd) integer.  After an
infinitesimal time interval $dt$, the sites will be occupied by $\phi$
and $A$, respectively, at the rate $a_{+}w_{10}(01)dt$
($a_{-}w_{10}(01)dt$).  We restrict the rates further as \BEA
&&w_{11}(10)=w_{11}(01)=w_{01}(10)=w_{10}(01)=1, \NN \\ 
&&w_{01}(11)=w_{10}(11)=\eta^{2}-1, \label{eq:pchoi} \EEA which makes
it possible to solve the model exactly by the IPDF method as will be
shown later.  Because we can fix the normalization of $a_{+}$ and
$a_{-}$ as $a_{+}+a_{-}=1$ by rescaling time, we put
$a_{\pm}=(1\pm\Delta)/2$ with the same $\Delta$ as in
(\ref{eq:spinhmt}).  The basic quantity in the stochastic model should
be the probability distribution function $P(\underline{\sigma};t)$,
which is the probability to find the system in the configuration
$\underline{\sigma}= (\sigma_{1},\cdots,\sigma_{L})$ at time $t$.
Defining the Hamiltonian by \BEA {\hat
  H}&=&\sum_{j=1}^{L}a_{(-)^{j}}{\hat H}_{j,j+1}\quad {\rm with}\quad
a_{(-)^{\rm odd}}=a_{-},\: a_{(-)^{\rm even}}=a_{+}, \\ 
\hspace{-23mm} \LR {\hat
  H}_{i,i+1}\RR^{\rho_{i}\rho_{i+1}}_{\sigma_{i}\sigma_{i+1}} &=& \LC
\begin{array}{lll}
  w_{\sigma_{i}-\rho_{i},\sigma_{i+1}-\rho_{i+1}}(\sigma_{i},\sigma_{i+1}),
  &{\rm if}&(\rho_{i},\rho_{i+1})\neq(\sigma_{i},\sigma_{i+1}) \\ 
  -\sum_{\alpha\neq 0,\beta\neq 0}
  w_{\alpha\beta}(\sigma_{i},\sigma_{i+1}), &{\rm
    if}&(\rho_{i},\rho_{i+1})=(\sigma_{i},\sigma_{i+1})
\end{array}
\right.  \EEA with periodic boundary conditions $i\equiv i+L$, we
can write the master equation of the probability distribution function
in terms of a Schr{\"o}dinger equation with imaginary time \BEA
\TD{}P(\sigma_{1},\cdots,\sigma_{L};t)&=&-\sum_{j=1}^{L}
\sum_{\rho_{j},\rho_{j+1}=1,0} a_{(-)^{j}}\LR {\hat H}_{j,j+1}
\RR^{\rho_{j}\rho_{j+1}}_{\sigma_{j}\sigma_{j+1}}\times\NN\\ &&\times
P(\sigma_{1},\cdots,\sigma_{j-1},\rho_{j},\rho_{j+1}
,\sigma_{j+2},\cdots,\sigma_{L};t).
\label{eq:master}
\EEA The Hamiltonian above given can be transformed into that of the
spin system (\ref{eq:spinhmt}) by the same mapping proceeded by
\cite{ann,bonn}.  Performing a similarity transformation ${\hat
  H}_{1}=U^{-1}{\hat H}U$ with the matrix \BE U=\LR\begin{array}{cc}
  \sqrt{\eta^{2}-1}&0 \\ 0&1
\end{array}\RR
\otimes\cdots\otimes \LR\begin{array}{cc} \sqrt{\eta^{2}-1}&0 \\ 0&1
\end{array}\RR,
\EE and a rotational transformation $H=R^{-1}{\hat H}_{1}R$ with \BE
R=\exp(i\theta s_{1}^{y}) \otimes\cdots\otimes \exp(i\theta s_{L}^{y})
\qquad (\tan\theta=\sqrt{\eta^{2}-1}), \EE we obtain the spin
Hamiltonian (\ref{eq:spinhmt}).

To calculate the correlation function from the master equation
(\ref{eq:master}), we use the IPDF method\cite{ann,bonn,doering3}
-\cite{henkel}.  For simplicity, we assume a Gaussian initial
condition for general $N$ and independent of $M$ as \BE
\sum_{\underline{\sigma}}P(\underline{\sigma};t=0)
\delta_{\sigma_{M},0}\cdots\delta_{\sigma_{M+N-1},0}
=p^{N},\label{eq:gauss} \EE where $p$ ($0\leq p\leq 1$) is the
probability of a site to be empty.  If we define the ``hole length
probability" (HLP) $\Omega(m;t)$, which is the probability to find
a string of empty sites with the length $m$ at time $t$, 
by 
\BEA \Omega(2n+1;t)&=&\sum_{\underline{\sigma}}
P(\underline{\sigma};t)
\delta_{\sigma_{M},0}\cdots\delta_{\sigma_{M+2n},0}, \NN \\ 
\Omega^{(\pm)}(2n;t)&=&\sum_{\underline{\sigma}}
P(\underline{\sigma};t)
\delta_{\sigma_{M},0}\cdots\delta_{\sigma_{M+2n-1},0},\quad (M={\rm
  even/odd}), \EEA we can easily verify that the above $\Omega$'s do
not depend on the starting site $M$ explicitly because of the
translational invariant initial condition (\ref{eq:gauss}). Note that 
there are two kinds of the HLP's
$\Omega^{(\pm)}$ for even hole-lengths, depending on whether the
starting point $M$ is even or odd.  With these preparations, we can
rewrite the master equation simply as \BEA \hspace{-23mm}
\TD{}\Omega(2n-1;t)&=& \eta^{2}\LS
a_{+}\Omega^{(+)}(2n;t)+a_{-}\Omega^{(-)}(2n;t) \RS
-(1+\eta^{2})\Omega(2n-1) \NN \\ \hspace{-23mm} &+&\LS
a_{+}\Omega^{(+)}(2n-2;t)+a_{-}\Omega^{(-)}(2n-2;t) \RS \quad{\rm
  for}~1\leq n\leq\HALF{L}, \NN \\ \hspace{-23mm}
\TD{}\Omega^{(\pm)}(2n;t)&=&2a_{\mp}\LS\eta^{2}\Omega(2n-1;t)
-(1+\eta^{2})\Omega^{(\pm)}(2n;t) +\Omega(2n+1;t)\RS \NN \\ 
&&\hspace{70mm}\quad{\rm for}~1\leq n\leq\HALF{L}-1. \label{eq:IPDF}
\EEA

The solution of Eq.(\ref{eq:IPDF}) is given by \BEA\hspace{-23mm}
\Omega(2n-1;t)\CS{=}\sum_{l=1}^{L/2-1}\sum_{s=\pm}
A_{l}^{(s)}\eta^{-(2n-1)} e^{\lambda_{l}^{(s)}t}\SIN{{\pi(2n-1)\over
    L}l}+ \NN \\ &&+B_{2n-1}e^{-(\eta^{2}+1)t}+\psi(2n-1), \NN\\ 
\hspace{-23mm} \OPM(2n;t)\CS{=}\sum_{l=1}^{L/2-1}\sum_{s=\pm}
A_{l}^{(s)}\eta^{-(2n+1)}{\lambda_{l}^{(s)}+2a_{\pm}(\eta^{2}+1) \over
  4a_{\pm}} e^{\lambda_{l}^{(s)}t}{\SIN{2\pi nl/L}\over\cos(\pi l/L)}+
\NN\\ &&+{1-a_{\pm}\over 1-2a_{\pm}}{2\over\eta^{2}+1}
(B_{2n-1}+\eta^{2}B_{2n+1})e^{-(\eta^{2}+1)t}+\psi(2n).\label{eq:solsym}
\EEA In the above expression, $A_{l}^{(\pm)}$ and $B_{2n-1}$ are
complicated coefficients depending on $L$, $\Delta$, $p$, $\eta$ and
$l$ and the zero-mode function is \BE \psi(m)={1\over 1-\eta^{-2L}}
\LS (1-p^{L})\eta^{-2m}+p^{L}-\eta^{-2L} \RS.
\label{eq:zeromode}
\EE The {\it energy spectrum} $\lambda_{l}^{(\pm)}$ takes the same
form as that in (\ref{eq:jwspectrum}).

With the above solution, we can investigate the finite size scaling of
physical quantities.  From now on, we set $\eta=1$, {\it i.e.}
the massless regime.  First, we consider the finite size correction of the
concentration $c(t)=1-\Omega(1;t)$ \cite{bonn}. It is not difficult to
ensure that we can fix a parameter $z=2(1-\Delta^{2})t/L^{2}$ finite
in the scaling limit $L\rightarrow\infty$ and $t\rightarrow\infty$.  
After performing a modular
transformation $ z\rightarrow -1/z$, the asymptotic form of the
concentration reads 
\BEA
c(t)\CS{\sim}\sqrt{1\over
    2\pi(1-\Delta^{2})t}\times \NN \\ 
\CS{\times} 
\LR 1-{1\over
    32(1-\Delta^{2})t} 
\LS {1+6p+p^{2}\over (1-p)^{2}}+
\Delta{3-2p+3p^{2}\over (1+p)^{2}}
\RS \RR +O(t^{-5/2}).  
\EEA

The finite size analysis for the two hole probability $\OPM(2;t)$,
which is identified with the probability to find a dimer on a link for
$a_{\pm}$, can be done similarly.  Due to alternation, $\OP(2;t)$  
and $\OM(2;t)$ do not coincide for general $t$ in spite of the
initial condition, $\OP(2;0)=\OM(2;0)=p^{2}$. However, because of the
diffusion and coagulation, we will observe 
$\OP(2;\infty)=\OM(2;\infty)=1$, which indicates the recovery of the
translational invariance. For example, let us consider 
the ratio of the two probabilities $\OP(2;t)$ and $\OM(2;t)$. 
After a lengthy calculation, its finite size scaling is
shown to be 
\BE {\OM(2;t)\over\OP(2;t)}\sim
1+{\Delta\over\sqrt{2\pi}} \LR(1-\Delta^{2})t\RR^{-3/2}+O(t^{-5/2}). 
\label{eq:findif}
\EE 
%

As we have seen, both the Jordan-Wigner method and the IPDF approach
are available in the XY model with diffusion and alternation.  An
energy gap proportional to the strength of the alternation is
observed.  We have also seen that the translational invariance is
recovered by the diffusion beginning with a translational invariant
initial condition. This recovery, {\it e.g.} $\OM(2;t)/\OP(2;t)$, is
scaled by the same variable $z$ as that of the concentration.  We have
obtained the exponents of the concentration and the two hole
probabilities in the scaling limit $L,t\rightarrow\infty$.

We can imagine some extension in this field. First, 
the coupling constants do not have to change on every other link  
in our approaches.  
That is, it is possible to change the coupling constant on every 
third (fourth and so on) site without loss of solvability. 
In particular, it would be interesting to consider the relation to  
the model with 
randomly changing coupling constants \cite{evans}. 
Secondly, contrary to the IPDF 
method, the Jordan-Wigner method can be also applied for the (partially) 
antiferromagnetic alternating model. Although we did not 
succeed in finding the 
corresponding initial condition in the Jordan-Wigner approach so far, 
we expect that a parallel discussion can be done. 

\section*{\small{\bf Acknowledgements}}
{\small 
The author would like to thank J.~Gruneberg,  
T.~Heinzel, Y.U.~Lee, R.~Raupach and V.~Rittenberg for reading 
the manuscript, 
helpful comments and suggestion. 
The discussions with J.~Suzuki and the theory group in the 
University of Bonn are also 
acknowledged. 
This work is financially supported by the AvH foundation. }


\begin{thebibliography}{99}
\bibitem{kadanoff}Kadanoff L P and Swift J 1968 \PR{\bf 165}~310
%
\bibitem{ann} Alcaraz F C, Droz M, Henkel M and Rittenberg V 1994 
\APNY{\bf 230}~250
%
\bibitem{pincus} Pincus P 1971 \SSC{\bf 9}~1971
%
\bibitem{okamoto} Okamoto K 1988 \JPSJ{\bf 57}~2947;~1990 \JPSJ{\bf 59}~4286; \\
Saika Y and Okamoto K 1995 "{\it Dimer} {\it correlation} {\it in} 
{\it spin-1/2} {\it alternative} XY {\it chain}" \\
e-print:~cond-mat/9510114
%
\bibitem{bonn} Hinrichsen H, Krebs K, Pfannm{\"u}ller M and Wehefritz B 
1995 \JSP{78}~1429
%
\bibitem{barouch} Barouch E,~McCoy B M~and~Dresden M 1970 \PR{\bf A2}~1075
%
\bibitem{hinrichsen}Hinrichsen H 1994 \JP{\bf A27}~1121
%
\bibitem{doering3} ben-Avraham D, Burschka M A and Doering C R 1990 
\JSP{60}~695
%
\bibitem{peschel}Peschel I, Rittenberg V and Schultze U 1994 
\NP{\bf B430}~633; \\ 
Sch{\"u}tz G 1995 \JSP{79}~243
%
\bibitem{henkel} Henkel M and Sch{\"u}tz G 1994 \PHYSICA{A206} 187; \\
Lushnikov A A 1986 \JETP{64}~811;~1987\PL{\bf A120}~135
\bibitem{evans} Evans M R 1996 \EPL{36}~13 
\end{thebibliography}
\end{document}